\documentclass[twoside,12pt]{article}
\usepackage{epsfig}
\usepackage{amsmath}

\def\Journal#1#2#3#4{{#1} {#2} (#4) #3 }

\def\PRD{{\em Phys. Rev.} D}

\def\INT{{\em Int. J. Mod. Phys.} A}

\def\p{\vec p}

\def\q{\vec q}

\newcommand{\be}{\begin{equation}}
\newcommand{\ee}{\end{equation}}
\newcommand{\bea}{\begin{eqnarray}}
\newcommand{\eea}{\end{eqnarray}}

\begin{document}

\title{ \vspace{1cm} Studying pion effects on the chiral phase transition}
\author{J.A. M\"uller$^{1,}\footnote{E-mail: jens.mueller@physik.tu-darmstadt.de\newline This work has been supported by a Helmholtz-University Young Investigator Grant No. VH-NG-332.}$\;,  
C.S. Fischer$^{1,2}$\\
\\
$^1$ Institute for Nuclear Physics, Darmstadt University of Technology, Germany\\
$^2$ GSI Helmholtzzentrum f\"ur Schwerionenforschung GmbH, Darmstadt, Germany}
\maketitle
\begin{abstract} 
We investigate the chiral phase transition at finite temperatures 
and zero chemical potential with Dyson-Schwinger equations. Our truncation 
for the quark-gluon interaction includes mesonic degrees of freedom,
which allows us to study the impact of the pions on the nature of the phase 
transition. Within the present scheme we find a five percent change of the 
critical temperature due to the pion backreaction whereas the mean field 
character of the transition is not changed.
\end{abstract}
%\eject
%\tableofcontents
%\section{Introduction and Quark DSE}
We investigate the chiral transition of two flavor QCD in the chiral limit 
(i.e. massless quarks). Assuming that effective restoration of axial vector 
symmetry takes place at temperatures above the critical temperature of chiral 
symmetry restoration the transition is expected to be a second order phase 
transition falling in the O(4)-universality class \cite{piswil0}. This motivates the application of effective models constructed in terms of order parameters to describe chiral transitions, see e.g. \cite{piswil0,boschwa0} and references therein. Explicit symmetry breaking changes the second order transition to a smooth 
crossover.

\begin{figure}[t]
\begin{center}
\epsfig{file=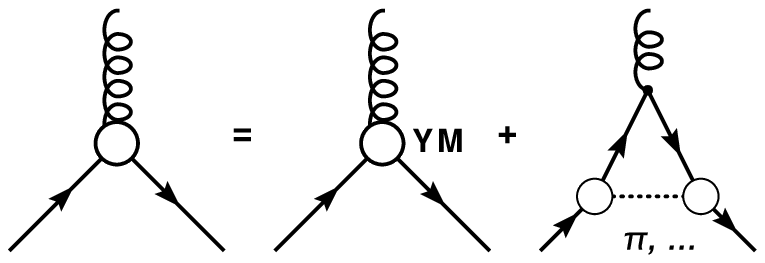,scale=0.8}\hspace{1.5cm} 
\epsfig{file=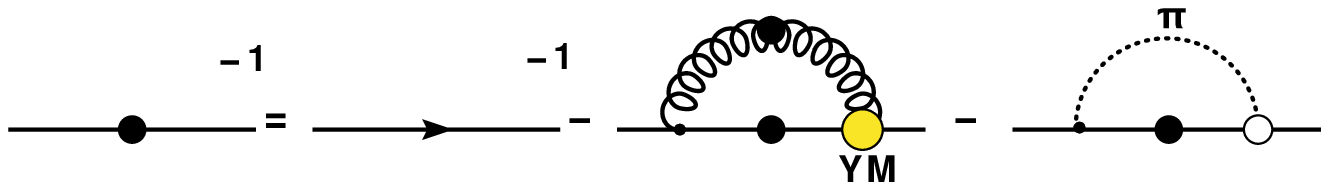,scale=0.8}
\end{center}
\caption{Approximated Dyson-Schwinger equations for the 
quark-gluon vertex and quark propagator.\label{fig1}}
\begin{center}
\epsfig{file=log_B_ohne_mit_pion2.eps,scale=0.35}
\end{center}
\caption{Scaling behaviour of $\chi_B:=B(\omega_0,\p^2=0)$ as a function of the
reduced temperature.  \label{fig3}}
\end{figure}

Here we study the chiral symmetry restoration with Dyson-Schwinger equations (DSE). 
In order to take into account the relevant degrees of freedom, i.e. the ones which 
retain long-range fluctuations in the vicinity of a second order phase transition, 
we need to employ a truncation of the quark-gluon vertex that includes mesonic 
fluctuations. Such a scheme has been proposed in \cite{finiwa0}. At 
finite temperatures it leads to a renormalised quark DSE of the form
\be
S^{-1}(\omega_n,\p)=Z_2 S_0^{-1}(\omega_n,\p) + Z_{1F}\; g^2\; \dfrac{4}{3}\; 
T \sum_{m=-\infty}^{\infty} \int \frac{d^3 q}{(2\pi)^3}\gamma_{\mu}
S(\omega_m,\q)\Gamma_{\nu}(\omega_m,\q;\omega_n,\p)D_{\nu\mu}(\Omega_{n-m},\p-\q),
\ee
where 
$S^{-1}(\omega_n,\p) = (\textrm{\textit{i}}\;
\vec{\gamma}\cdot \p\;A(\omega_n,\p)+\textrm{\textit{i}}\;
\gamma_4 \omega_n\;C(\omega_n,\p)+B(\omega_n),\p)$ 
is the inverse dressed quark propagator with dressing functions 
$A,\;B,\;C$ and $S_0^{-1}$ its bare counterpart. $Z_2$ and $Z_{1F}$ are
renormalisation constants and $\omega_n=\pi T (2n+1)$ is the fermion 
Matsubara frequency. In the quark self energy we have the dressed gluon 
propagator $D_{\nu\mu}$ and the quark-gluon vertex $\Gamma_{\nu}$. The truncation
of the quark-gluon vertex together with the resulting form of the quark-DSE is given
diagrammatically in Fig.\ref{fig1}. The symbol 'YM' denotes contributions from the
gluonic sector of QCD whereas the meson exchange diagram contains in principle 
all possible mesonic fluctuations, see \cite{finiwa0} for details. 
As a first step towards a more complete description, the preliminary results reported 
here include only pion contributions on a qualitative level.

%\section{Results}
One possibility to characterise a phase transition is to study order parameters.
For a second order phase transition there is a scaling region where the chiral
condensate $\langle \bar{\psi} \psi \rangle_{\mu}$ is proportional to $\sim t^{\beta}$ with $t=(1-T/T_c)$ 
the reduced temperature and $\beta$ the critical exponent. We found numerically that in the 
 chiral limit $\chi_B:=B(\omega_0,\p^2=0)$ equally serves as an order parameter since for $t\sim 0$: $\chi_B\propto \langle \bar{\psi} \psi \rangle_{\mu} \propto t^{\beta}$. To extract this behaviour from numerical calculations the critical temperature has to be determined to some accuracy since the scaling sets in not until $t < 0.03$ and the slope in the vicinity of small $t$ is strongly 
dependent on deviations from $T_c$. In our (qualitative) study the inclusion 
of the pion loop decreases the critical temperature by $\sim 5$\% 
from $T_c\sim 199$ MeV to $T_c \sim 188$ MeV. The scaling behaviour
of $\chi_B$ is shown in Fig.\ref{fig3}. Both our results with and without the 
pion backreaction can well be fitted by a power law $\sim t^{1/2}$, which is characteristic
of mean field behaviour. (For the vertex dressing without the pion this is in 
accordance with previous results in the DSE approach summarised in 
\cite{Roberts:2000aa}.) These results suggest that the 
contributions necessary to go beyond mean field are not yet included in our approach.
So far we do not take care of all temperature dependencies of the pion wave function 
and decay constants. Furthermore we neglected contributions from the scalar meson channel.
A detailed study of these effects is under way and will be reported elsewhere.

\end{document}